# Nucleosynthesis of "Light" Heavy Nuclei in Neutrino-driven Winds. Role of (α,n) reactions

Jorge Pereira[1,2], Almudena Arcones[3,4], Julia Bliss[3] and Fernando Montes[1,2]

[1] National Superconducting Cyclotron Laboratory, Michigan State University, E. Lansing, MI 48824, USA
[2] Joint Institute for Nuclear Astrophysics, Michigan State University, E. Lansing, MI 48824, USA
[3] Institut für Kernphysik, Technische Universität Darmstadt, Schlossgartenstr. 2, Darmstadt, D-64289, Germany
[4] GSI Helmholtzzentrum für Schwerionenforschhung GmbH, Planckstr. 1, Darmstadt, D-64291, Germany

E-mail: pereira@nscl.msu.edu

## Abstract

Neutrino-driven winds following core collapse supernovae have been proposed as a suitable site where the so-called light heavy elements (between Sr to Ag) can be synthetized. For moderately neutron-rich winds, (α,n) reactions play a critical role in the weak r process, becoming the main mechanism to drive nuclear matter towards heavier elements. In this paper we summarize the sensitivity of network-calculated abundances to the astrophysical conditions, and to uncertainties in the (α,n) reaction rates. A list of few (α,n) reactions were identified to dominate the uncertainty in the calculated elemental abundances. Measurements of these reactions will allow to identify the astrophysical conditions of the weak r process by comparing calculated/observed abundances in r-limited stars.

## 1. Introduction

One of the most remarkable features of the r process is the nearly perfect match of the elemental abundance pattern (particularly in the rare-earth region) observed in r-process-rich stars (see, e.g. Fig 5 of [1]). These are halo, metal-poor [Fe/H] < -1 (old) stars, characterized by high levels of Eu enrichment (either [Eu/Fe] > +1, for r-II stars, or 0.3 ≤ [Eu/Fe] ≤ +1, for r-I stars [2]) and low levels of s-process "contamination" ([Ba/Eu] < 0). Strikingly, the same pattern is found in the solar-system (SS) r-process residual elemental abundances (obtained by subtracting the s- and p-process contributions to the observed abundances). This regularity points to a robust r process, typically referred to as main r process, operating over (at least) the age of the Galaxy.

A closer examination of these observed elemental abundances, however, reveals some interesting features in the region of lighter elements (38<Z<47). First, the abundances of these elements exhibit a notorious scatter between different r-process-rich stars, in contrast to the nearly perfect match observed for elements Z>56. Second, there seems to be a consistent depletion of light elements in these r-II stars (in particular Y, Mo, Rh, Pd, and Ag), compared with the Eu-scaled SS r-process residuals. Several authors investigated these anomalies by comparing the observed abundances of lighter heavy element with the Ba- and/or Eu-enrichments for different metal-poor stars (see e.g. [3-6]). As an example, Fig. 1 shows the abundances of a typical light heavy element (e.g. Sr), normalized to Eu, as a function of Eu/Fe (i.e., r-process enrichment) for a sample of Eu-rich metal-poor stars. There are two clearly distinct regions in this figure: First, stars with low levels of Eu-enrichment exhibit a linear anti-correlation between [Sr/Eu] and [Eu/Fe], indicating that the Sr and Eu "contaminations" in these stars were produced by different processes. Second, there is a clear correlation of the Sr and Eu abundances in highly Eu-rich stars, as seen from the flat trend of [Sr/Eu] vs. [Eu/Fe]. Note that this is the same correlation observed when analyzing the abundances of rare-earth elements normalized to Eu, as shown in Fig. 2 for [La/Eu].

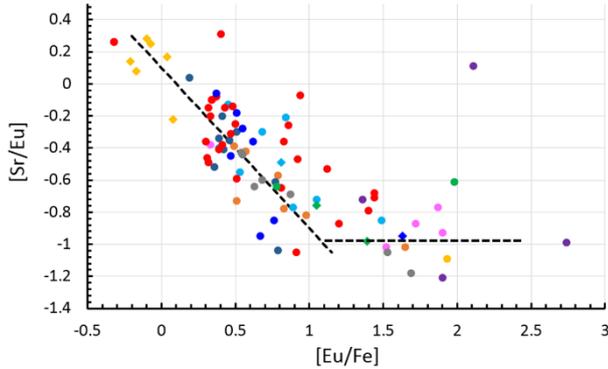

*Figure 1: [Sr/Eu] as a function of [Eu/Fe] for different samples of metal-poor halo stars (data taken from JINAbase [7]).*

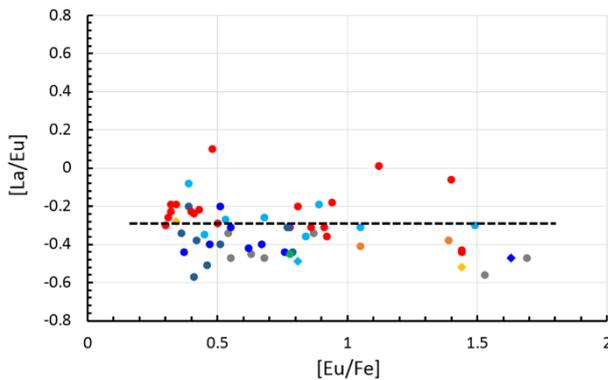

*Figure 2: [La/Eu] as a function of [Eu/Fe] for different samples of metal-poor halo stars (data taken from JINAbase [7]).*

The picture emerging from these results is that there seems to be (at least) a secondary process, besides the main r process, co-producing lighter heavy elements (38<Z<47), with no significant productions of heavier elements (Z>56). Thus, the SS r-process residuals might result from a combination of the main r process and this secondary process. In that sense, the element abundances of the so-called r-limited stars (i.e. [Eu/Fe] < 0.3; [Sr/Ba] > 0.5; [Sr/Eu] > 0 [2]) might be offering the observational signature of this secondary process. This is illustrated in Fig. 3, where the element abundance pattern of one of these r-limited stars (HD122563) is compared with the Eu-scaled pattern of a typical r-II star (CS22892-052).

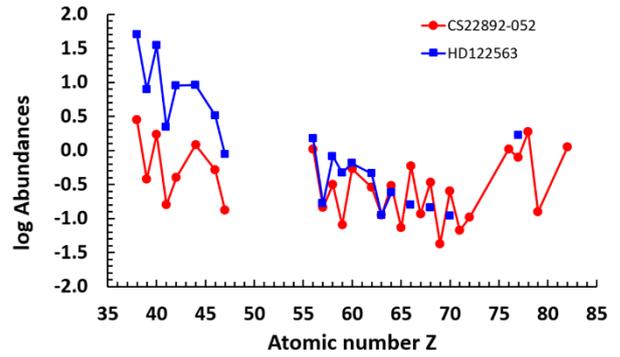

*Figure 3: Eu-normalized elemental abundances of HD122563 and CS22892-052 stars (data taken from JINAbase [7]).*

In this paper, we discuss the nucleosynthesis process occurring in slightly neutron-rich neutrino-driven winds produced in core-collapse supernovae (CCSNe). A brief description of this process is given in Section 2. Section 3 concerns the sensitivity of this process to the astrophysical conditions found in CCSNe neutrino-drive winds. The role of nuclear-physics uncertainties in the network-calculated abundances is discussed in Section 4. Summary and conclusions are included in Section 5.

## 2. Nucleosynthesis in Core-Collapse Supernova Neutrino-driven Winds: The role of ($\alpha$,n) reactions

Neutrino-driven winds following CCSNe were originally considered as an attractive site for the nucleosynthesis of the r process. However, after years of intense effort, the current status is that hydrodynamic simulations do not reach the extreme conditions (in terms of entropy S, expansion times $t_{exp}$, and electron fractions $Y_e$) necessary for a robust nucleosynthesis process capable of reproducing the main r-process pattern (see section 3.1.1 of [8] for an interesting historical overview). Nevertheless, the moderate neutron-rich (in some cases even proton-rich) wind conditions obtained from these simulations offer an optimum environment for the synthesis of the lighter heavy nuclei [9]. In particular, in fast-expanding ejecta, the environment remains slightly neutron-rich ($Y_e \lesssim 0.5$) and it is possible to have a weak r process that "fails" to produce nuclei beyond A~110 (see, e.g. [10]).

As discussed in [11,12], near the proto-neutron star, the hot ejected matter is dominated by neutrons and protons that form alpha particles in nuclear statistical equilibrium (NSE). As this matter expands and its temperature and density drop, alpha particles recombine into heavier nuclei via the slow $3\alpha$ reaction, combined with $\alpha(\alpha,n)^9\text{Be}(\alpha,n)^{12}\text{C}$, and other reactions, including (n,$\gamma$)–($\gamma$,n), in NSE. During this NSE phase, a series of "seed" nuclei are created, which can reach elements in the region around Zn. If the expansion is very fast ($t_{exp}$~10 ms), the abrupt drop in temperature leads to an alpha-rich NSE freeze-out at T~5 GK. In these conditions, ($\alpha$,n) reactions are the fastest, out-of-equilibrium, reactions pushing matter towards heavier elements, with minor contributions from (p,n), ($\alpha$,$\gamma$) and (p,$\gamma$) reactions. Meanwhile, for each new Z, the equilibrated (n,$\gamma$)–($\gamma$,n) reaction sequences determine the isotopic matter distribution. Note that at these moderate neutron-rich conditions, the freshly synthesized matter follows a path close to stability, in the neutron-rich side. This charged-particle reaction phase (CPR), also referred to as alpha process, lasts until matter cools down below T~2 GK. Below this temperature, there is a charged-particle freeze-out, followed by a sequence of (n,$\gamma$)–($\gamma$,n) and beta decays until neutrons are exhausted and all matter decays to the valley of stability, leading to the final observable abundance distribution.

As discussed in our previous papers [12-14], reaction-network calculated abundances produced in this weak r-process are affected by two sources of uncertainty: First, the astrophysical conditions of the environment ($Y_e$, entropy, and expansion time) depend on uncertainties in neutrino interactions, hydrodynamics, explosion energies, etc. Second, of all the ($\alpha$,n) reactions involved in the process, none of them are experimentally known in the relevant range of temperatures (T~2-5 GK). In the following sections, we discuss the sensitivity of this weak r process to the astrophysical environment and to the ($\alpha$,n) rate uncertainties.

## 3. Sensitivity of the Lighter Heavy-element production to Astrophysical Conditions

The sensitivity of the weak r-process nucleosynthesis to the astrophysical conditions of the neutrino winds were extensively investigated in one of our recent article [14]. This study followed a two-step method. In the first step, the neutrino-wind properties were determined using the steady-state model of Otsuki et al. [15], which relies on the fact that in the first seconds following the core collapse, the proto-neutron star mass and radius, and the neutrino properties (luminosity and energy) vary slowly in time. Steady-state wind equations can then be used to calculate the pressure, temperature, velocity, and density as functions of the radius r (defined as the distance to the center of the proto-neutron star), for a given mass $M_{NS}$, radius $R_{NS}$, and (anti)neutrino luminosities (L) and energies ($\epsilon$). As explained in [14], the range of input $M_{NS}$ and $R_{NS}$ considered in our study (0.8-2 $M_\odot$ and 9-30 km) were chosen based on observational and theoretical constrains for neutron stars and neutron matter. Moreover, the (anti)neutrino properties were selected to cover a range of moderately neutron-rich winds (i.e. $Y_e$=0.40-0.49).

Thus, for each set of input parameters, we produced a "wind trajectory", characterized by the electron fraction $Y_e$ and the evolution of the temperature T(t) and density $\rho$(t), after converting the velocity as a function of r into time. These trajectories are associated with a set of wind parameters given by S, $Y_e$, and $t_{exp}$[1].

In the second step of our method, each of the calculated wind trajectories were used in a reaction network to determine the production of nuclei associated with these astrophysical conditions. The ($\alpha$,n) reaction rates included in this network were calculated with the TALYS code [16], using the default set of input parameters defined in [13], except for masses, which were taken from the 2003 mass compilation of Audi *et al.* [17] (we will refer to these TALYS rates as TALYS1).

Following this two-step method, we selected 2696 different sets of input wind parameters ($M_{NS}$, $R_{NS}$, $Y_e$), and produced, for each of them, a different set of abundances. Remarkably, all of these nearly 3000 abundance sets can be grouped in just four clearly-different abundance patterns, labeled NSE1, NSE2, CPR1, and CPR2, as shown in Fig. 4. More importantly, the underlying astrophysical conditions of each of these four patterns can be reduced to two quantities, namely, the neutron-to-seed ratio ($Y_n/Y_{seed}$) and the alpha-to-seed ratio ($Y_\alpha/Y_{seed}$) at temperatures T≈3 GK (see Fig. 5).

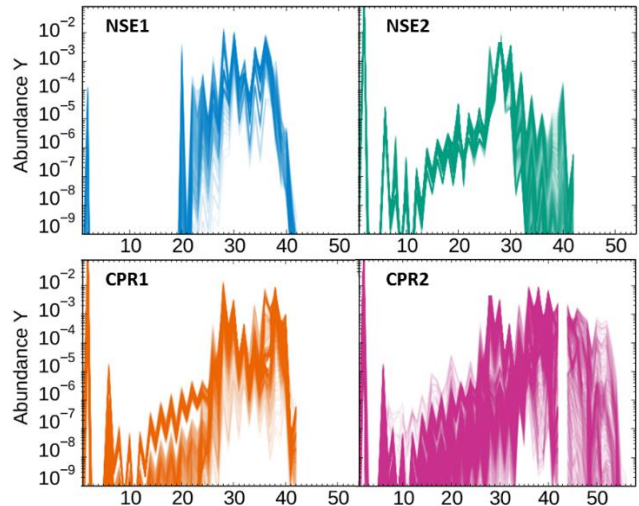

*Figure 4: The four characteristic element abundance patterns obtained in [14] using different astrophysical conditions.*

---

[1] Note that S ~ $T^3/\rho$ and $t_{exp}$=r/v at T=0.5 MeV

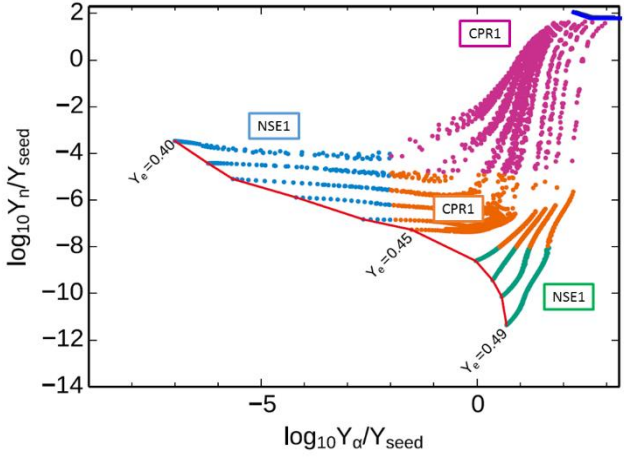

*Figure 5: The four abundance pattern groups (NSE1, NSE2, CPR1, CPR2) of [14] in the $Y_\alpha/Y_{seed}$ vs. $Y_n/Y_{seed}$ plane. The blue and red lines correspond to the most and least compact proto-neutron stars (see [14] for more details).*

As explained in [14], of the four identified abundance patterns shown in Fig 4, the group labeled CPR2, resulting from the most compact proto-neutron stars with the largest $Y_n/Y_{seed}$ and $Y_\alpha/Y_{seed}$ values, is the only one with significant productions of elements between Sr and Ag. This is the result of two combined effects: First, because the neutron-to-seed ratio is large, the seed nuclei produced during NSE are localized in the neutron-rich side with maximum productions in the N=50 closed shell for elements between Se and Kr. Second, the high alpha-to-seed ratio values at the end of NSE guarantee that there are enough alpha particles that can be captured in ($\alpha$,n) reactions pushing the seed nuclei towards heavier elements through the N=50 closed shell. These ($\alpha$,n) combined with (n,$\gamma$)–($\gamma$,n) give rise to a successful CPR producing large amounts of neutron-rich isotopes, which will eventually beta-decay to stability, leading to the final abundance pattern.

## 4. Sensitivity of the Lighter Heavy-element production to ($\alpha$,n) reaction-rate uncertainties

As discussed in the previous sections, ($\alpha$,n) reactions are by far the most important mechanism transmuting the NSE seed nuclei into heavier elements, and are consequentially expected to have a strong impact in the final abundances. Interestingly, almost none of the approximately 900 different ($\alpha$,n) reactions involved in our network calculations have been measured in the range of temperatures T≈1-5 GK, relevant for the weak r process. (Note that, unlike the main r-process, other important nuclear-physics inputs like masses or half lives are well known experimentally.). Thus, one relies on global reaction codes like e.g. TALYS to determine these reaction rates. As discussed in [12], comparisons of TALYS-calculated ($\alpha$,n) reaction rates with values measured at temperature above 5 GK show discrepancies in the order of a factor 10. At lower temperatures, the uncertainty of the reaction codes must be theoretically evaluated. As discussed by Pereira *et al* [13] (see also [14]), the main source of uncertainty of the calculated ($\alpha$,n) reaction rates is the alpha optical potential, which can lead to differences as high as two orders of magnitude, depending on the temperature considered. Moreover, the most important ($\alpha$,×n) channel in the weak r process is ($\alpha$,1n).

Taking into account these uncertainties, we have recently investigated the impact of the ($\alpha$,n) reaction rates on the calculated weak r-process abundances using a Monte-Carlo sampling method [19]. In order to take into account the uncertainties in the astrophysical conditions discussed in Section 3, we performed this sensitivity study for 35 different trajectories selected within the hundreds of different trajectories describing the CPR2 group (see Fig. 5). This method, discussed in detail in [19], followed several steps: First, we selected one of the 35 wind trajectories representing the CPR2 group. Then, each of the ($\alpha$,n) reaction rates calculated with TALYS1 (see Section 3), were scaled by a unique randomly selected factor p, which was chosen following a log-normal distribution with a sigma value determined on the basis of the theoretical uncertainties discussed above (here p=1 corresponds to the original rates obtained from TALYS1). The new "scaled" set of nearly 900 ($\alpha$,n) reaction rates was then used in our reaction network to calculate the final abundances. This sequence was repeated 10,000 times for each of the 35 astrophysical trajectories considered. Preliminary results from this study are shown in Fig. 6, where the lighter heavy-element element abundances were calculated under different assumptions: First, the elemental-abundance pattern obtained with TALYS1 (i.e., scaling factor p=1) is shown (black thick solid line) for one of the 35 selected astrophysical trajectories (labeled MC6). Second, variations of these elemental abundances due to variations of the ($\alpha$,n) rates (i.e. p scaling factor randomly distributed) are shown by the dark-magenta region. As can be seen, uncertainties in the ($\alpha$,n) rates can lead to uncertainties in the calculated abundances as high as two orders of magnitude. Finally, the light-orange lines obtained with TALYS1 (p=1) for each of the different CPR2 trajectories illustrate the variation in the calculated abundances entirely due to uncertainties in the astrophysical conditions.

The Zr-scaled elemental abundances of the so-called Honda Star HD122563 are included in Fig. 6. As can be seen, our calculations reproduce these abundances, within the ($\alpha$,n)-uncertainties, for the selected astrophysical trajectory MC6 (black line). From this result, one might be tempted to conclude that the astrophysical conditions of the selected trajectory MC6 are the right ones for a successful weak r process (i.e. a process capable of reproducing the abundances of Sr-Ag element observed in HD122563).

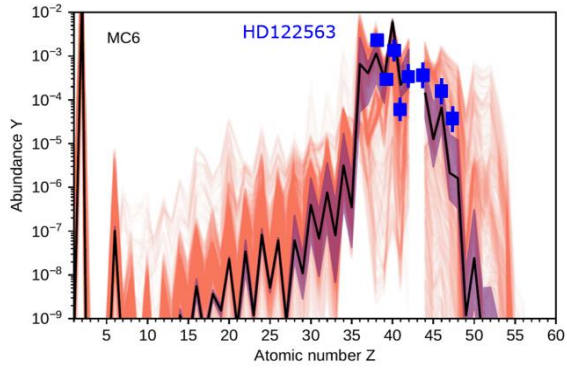

*Figure 6: Elemental abundances obtained in the weak r process. Solid black line corresponds to abundances obtained using the TALYS1 code and one of the 35 CPR2 astrophysical trajectories (MC6). Dark-magenta region corresponds to variations of these elemental abundances due to variations of the ($\alpha$,n) rates. Light-orange region corresponds to variations of the calculated abundances due to astrophysical uncertainties. Blue squares show the Eu-scaled abundances of the r-limited HD122563 star.*

However, this is not the only "successful" astrophysical condition. For instance, Fig. 7 shows how a different trajectory (labeled MC30) can also reproduce the abundances of HD122563, within the ($\alpha$,n)-uncertainties.

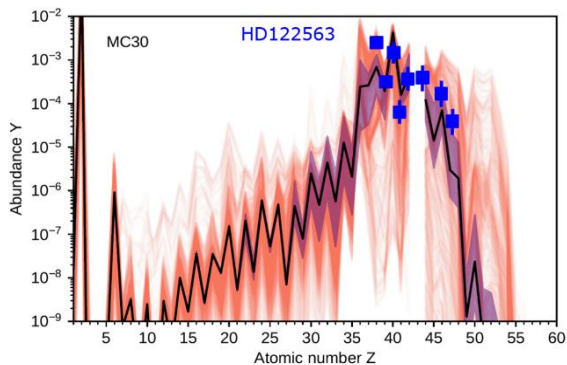

*Figure 7: Same as Fig. 6, using the trajectory MC30 instead of MC6.*

As shown in Figs. 6 and 7, in order to identify the actual astrophysical conditions of a successful weak r process, it is critical to reduce the theoretical uncertainty of the ($\alpha$,n) reaction rates by measuring the corresponding reaction cross sections at the relevant energies (temperatures). Since this is a cumbersome task, given the fact that there are more than 900 reactions affecting the abundances of elements in the region Z=36–47. An alternative approach is to identify which are the most important ($\alpha$,n) reactions that need to be measured. In order to answer this question, we analyzed the correlations between each of this ($\alpha$,n) reactions and the abundance of each of the elements produced in our network calculations. An example of this study is shown in Fig 8, which shows the abundance of Ag as a function of the scaling factor p applied to the reaction $^{94}$Sr($\alpha$,n)$^{97}$Zr.

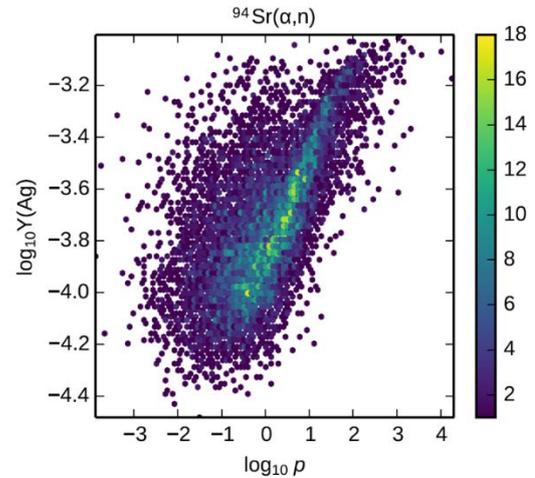

*Figure 8: Calculated Ag abundance as a function of the scaling factor p applied to the reaction $^{94}$Sr($\alpha$,n)$^{97}$Zr.*

As described in detail in [19], these correlation plots were used to identify the most important ($\alpha$,n) reactions that need to be measured. From this study, we obtained a total of 45 reactions, organized in four priority groups. The priority rank depends on how many elements are affected by the reaction, and for how many CPR2 astrophysical trajectories the reaction is important. Some examples of the reactions identified for each of these priority groups are:

- **Priority I** (reaction affecting many elements, under many different astrophysical conditions): $^{86-90}$Kr($\alpha$,n)$^{89-93}$Sr
- **Priority II** (reaction affecting few elements, under many different astrophysical conditions): $^{85,87}$Br($\alpha$,n)$^{88,90}$Rb
- **Priority III** (reaction affecting many elements, under few different astrophysical conditions): $^{72,76,78-80}$Zn($\alpha$,n)$^{75,79-83}$Ge
- **Priority IV** (reaction affecting few elements, under few different astrophysical conditions): $^{67,77}$Cu($\alpha$,n)$^{70,80}$Ga

The complete list of reactions, including the elements affected by them and the relevant astrophysical conditions will be presented in a forthcoming publication [19].

## 5. Conclusions

Several anomalies in the elemental abundances observed in metal-poor stars suggest the need of a secondary process, besides the so-called main r process, to explain the abundance pattern of light heavy elements (between Sr and Ag). This secondary pattern might be responsible for the abnormal overproduction of elements like Sr, compared to Eu, in the so-called r-limited stars.

Neutrino-driven winds, following core collapse supernovae, offer an interesting site where the nucleosynthesis of these elements might occur. For slightly neutron-rich winds, the so-called weak r process produces nuclei below the so-called second r-process peak, which includes elements like Sr, Y, and Zr.

A series of papers published in the last years by our collaboration, explored the role of (α,n) reactions in this weak r process. A steady-state model was used to determine the wind trajectories [$Y_e$, T(t), ρ(t)] from a large sample of proto-neutron star masses and radii, and (anti)neutrino properties. These trajectories were then coupled to a network calculation to determine the resulting abundances. The most noticeable aspects of that work are: (1): for moderate neutron-rich winds ($Y_e$=0.40–0.49), four clearly distinguishable abundance patterns were obtained; they were labeled as NPE1, NPE2, CPR1, and CPR2; (2) each of these patterns is correlated with the alpha-to-seed $Y_\alpha/Y_{seed}$ and neutron-to-seed $Y_n/Y_{seed}$ ratios at temperatures T≈3 MeV; (3) the pattern CPR2, corresponding to the most compact proto-neutron stars (leading to relatively high values of $Y_\alpha/Y_{seed}$ and $Y_n/Y_{seed}$) favor the synthesis of elements around Sr to Ag; (4) under these conditions, (α,n) reactions play a crucial role to synthesize heavier nuclei; (5) none of these (α,n) reactions are experimentally known; (6) differences in alpha optical potential models can lead to variation in the calculated (α,n) reaction rates up to two orders of magnitude.

Following these results, multiple reaction-network calculations of elemental abundances were performed assuming different astrophysical conditions within the CPR2 group. For each calculation, a randomly-selected scaling factor, reflecting the (α,n) rate uncertainty, was independently applied to each of the almost 900 different (α,n) reactions involved. The calculated light heavy-element abundances were found to vary by up to two orders of magnitude due to uncertainties in the (α,n) reaction rates. In order to identify the most important (α,n) reaction, the calculated abundance of each element, between Kr and Ag, were plotted as a function of the scaling factor applied to each. By analyzing the correlations between the abundance of each element and the scaling factor applied to each reaction, we identified a list of the most important (α,n) reactions. Measurements of these reactions will substantially reduce the uncertainty of the calculated abundances. This, in turn, will allow to identify the astrophysical conditions necessary for the weak r process by comparing our calculated abundances with observations in r-limited stars.

**Acknowledgements**

This work was funded by Deutsche Forschungsgemeinschaft through SFB 1245, ERC 677912 EUROPIUM, BMBF under grant No. 05P15RDFN1, and by the National Science Foundation under Grant No. PHY-1430152 (JINA Center for the Evolution of the Elements). J.B. acknowledge the MGK of the SFB 1245 and the JINA Center for the Evolution of the Elements for support during a research stay at Michigan State University